\documentclass[12pt,preprint]{aastex}

\usepackage{graphicx}

\newcommand{\HII}{{\sc H\,ii}}

\newcommand{\Halpha}{H$\alpha$}
\newcommand{\sigpol}{$\Sigma_{pol}$}

\shorttitle{Polarization shadows}
\shortauthors{Stil \& Taylor}

\begin{document}

\title{Polarization shadows of extragalactic sources by the local magneto-ionic ISM}

\author{J. M. Stil\altaffilmark{1} and A. R. Taylor}
\affil{Department of Physics and Astronomy, University of Calgary}
\altaffiltext{1}{stil@ras.ucalgary.ca}

\begin{abstract}
We report depolarization of extragalactic sources in the NRAO VLA Sky
Survey (NVSS) by local structures in the interstellar medium. The sky
density of polarized sources drops by a factor 2 -- 4 in regions with
angular scales $\sim 10\degr$, implying up to 40\% depolarization on
average per source.  Some of these polarization shadows are associated
with \HII\ regions, but three are associated with regions of
depolarized diffuse Galactic emission. The absence of a correlation
between the depth of polarization shadows and \Halpha\ intensity
suggests that some shadows are related to structure in the magnetic
field.  At least some polarization shadows are caused by partial
bandwidth depolarization in the NVSS.  Alternatively, some may be
caused by regions with small scale ($\lesssim 1\arcsec$) variations in
rotation measure.
\end{abstract}



\section{Introduction}

Observations of polarized radio emission provide a way to study
magnetic fields in galaxies. The observed plane of polarization of
optically thin, linearly polarized synchrotron emission is
perpendicular to the direction of the magnetic field projected on the
sky, unless Faraday rotation changes the plane of polarization as the
radiation travels through a plasma with a magnetic field component
along the line of sight.  The plane of polarization of emission with
wavelength $\lambda$ (m) rotates by an angle $\Delta \theta$ (rad),
according to
\begin{equation}
\Delta \theta = 0.81 \lambda^2 \int_{source}^{observer} n_e B_\| dl = \lambda^2 RM
\label{RM-eq}
\end{equation}
with electron density $n_e$ (cm$^{-3}$), line of sight component of
the magnetic field $B_\|$ ($\mu$G), and line of sight distance element
$dl$ (pc). The rotation measure $RM$ ($\rm rad\ m^{-2}$) is a property
of the medium along the line of sight from the source to the observer.
Rotation measures can be derived from multi-frequency observations of
polarized emission, fitting $\Delta \theta$ as a function of
$\lambda^2$.  Rotation measures of compact polarized sources have been
used to constrain models of the Galactic magnetic field,
e.g. \citet{han1999,brown2003}.

Observations of diffuse polarized Galactic emission have revealed
structures in the magneto-ionic medium that remain undetected in total
intensity, because an ionized region can have negligible emission
measure, yet still produce measurable Faraday rotation
\citep{uyaniker2003}.  However, interpretation of structures in
diffuse polarized emission is difficult, as both polarized emission
and Faraday rotation may occur intermixed anywhere along the line of
sight.

\citet{gaensler2005} and \citet{haverkorn2006} reported depolarization
of extragalactic sources by regions of high emission measure in the
LMC and the Southern Galactic Plane Survey \citep{haverkornetal2006b}
respectively.  In this Letter, we report detection of depolarization
of extragalactic sources on much larger angular scales by the local
magneto-ionic medium.

\section{Polarization shadows}

The main dataset for our analysis is the NRAO VLA Sky Survey (NVSS)
\citep{condon1998}, a single-frequency 1.4 GHz continuum survey of the
sky north of declination $-40\degr$ with the VLA in D configuration
($45\arcsec$ angular resolution). The NVSS bandwidth is the sum of two
42 MHz wide bands, but separated by 70 MHz in frequency.  The
principal data products are images in Stokes I, Q, and U, and a source
catalog of 2-dimensional Gaussians fitted to compact structures in
these images. The mean noise level of the NVSS in Q and U is 0.29 mJy
beam$^{-1}$.

Figure~\ref{shadows-fig} shows images of the number density \sigpol\
(sources per square degree) of NVSS sources with polarized intensity
$p > 1$ mJy beam$^{-1}$, smoothed to a resolution of $2\degr$,
\Halpha\ intensity composed by \citet{finkbeiner2003} from the VTSS
\citep{dennison1998}, SHASSA \citep{gaustad2001}, and WHAM
\citep{haffner2003} surveys, and diffuse polarized intensity at 1.4
GHz from the DRAO 26-m polarization survey \citep{wolleben2006}.  The
number density of polarized sources shows distinct depressions (dark
areas) on angular scales $\sim10\degr$, some of which coincide with
enhanced \Halpha\ emission. Most conspicuous is a ring centered on
$(l,b) = (-106\degr,+4\degr)$ that coincides with an \Halpha\ shell
associated with the Gum nebula.  We refer to these structures as
polarization shadows.

For the most significant features in Figure~\ref{shadows-fig}a,
Table~\ref{shadows-tab} lists Galactic coordinates, angular diameter,
minimum source density $\Sigma_{pol,min}$, \Halpha\ intensity averaged
over a 2 degree box at the minimum of \sigpol, identification with
known \HII\ regions, and distance. These features have values of
\sigpol\ a factor 2 -- 4 smaller than the mean background of 6.6
sources deg$^{-2}$.  The rms noise level $\sigma$ in
Figure~\ref{shadows-fig}a caused by statistical fluctuations in
\sigpol, is 1.3 sources deg$^{-2}$. The contour in
Figure~\ref{shadows-fig}a at half the background source density is
therefore 2.5$\sigma$ below the mean value of
\sigpol. Table~\ref{shadows-tab} is complete for features with
$\Sigma_{pol,min}< 2.7$ deg$^{-2}$. A comprehensive analysis of
smaller structures will be deferred to a following paper.

The association of several polarization shadows with nearby \HII\
regions is the strongest indication that these shadows are real. \HII\
regions are also known to depolarize diffuse emission
\citep{gray1999,gaensler2001}. We inspected NVSS images, and also
verified that the structures were robust against variation of the
threshold in $p$.  The polarization shadows have no counterpart in the
all-sky images of total NVSS source density, or NVSS noise
amplitude. The structures are also visible in an image of mean
fractional polarization, demonstrating that the mean fractional
polarization is lower in these regions.  Some apparent excesses in
\sigpol\ are seen near bright radio emission in the Galactic
plane. These excesses, seen in white in Figure~\ref{shadows-fig}a, are
the result of confusion from small-scale structure in Galactic
emission, and residual sidelobes around bright emission. We exclude
these regions from our analysis.

The NVSS contains small areas (typically $<1\degr$ across) with
missing polarization data. Most of these areas exist along lines of
constant right ascension in the declination range $-10\degr$ to
$+25\degr$, which intersects the Galactic equator at $l \approx
+40\degr$.  We have verified that the features in
Table~\ref{shadows-tab} are not the result of holes in the sky
coverage of the NVSS. We have identified these holes and their effect
on \sigpol\ is almost always lost in the noise.

Three extended polarization shadows listed in Table~\ref{shadows-tab}
are found in regions where $I_{H\alpha}\lesssim 10$ R. Despite the low
\Halpha\ intensity, these shadows are among the deepest in
$\Sigma_{pol,min}$.  These shadows coincide with minima in diffuse
polarized emission in the DRAO 26-m polarization survey
\citep{wolleben2006} shown in Figure~\ref{shadows-fig}c. One of these
is shown in Figure~\ref{shadows_detail-fig}. The features seen in the
DRAO survey also exist in the Effelsberg medium latitude survey by
\citet{uyaniker1999}.  This polarization shadow coincides with an edge
in diffuse polarized emission.  A diffuse polarized intensity filament
coincides with a higher polarized source density between the two
strongest shadows.  Large changes in polarization angle are also seen
associated with the depressions in \sigpol.

The polarization shadow in Figure~\ref{shadows_detail-fig} overlaps in
part with the Canadian Galactic Plane Survey
\citep{taylor2003}. \citet{uyaniker2003} note a diagonal band of low
diffuse polarized intensity extending from $(l,b) = (87\fdg6,
+4\fdg5)$ to $(l,b) = (94\fdg0, -2\fdg0)$, the approximate course of
the polarization shadow. A triangular region of depolarization noted
by these authors at $84\fdg5 < l < 89\fdg5$, $b = -3\degr$ coincides
with the deeper part of this shadow. Rotation measures of polarized
sources in this region are $\sim -350$ rad m$^{-2}$, with large
variations on the scale of a few degrees
\citep{brown2003}. Unpublished rotation measure data with more
complete longitude coverage confirm a negative excess in rotation
measure in the longitude range $84\degr < l < 90\degr$ (Jo-Anne Brown,
private communication) that appears to be associated with this
polarization shadow.

\section{Discussion}

Polarization shadows in Figure~\ref{shadows-fig} indicate
depolarization of the background sources, leaving fewer sources with
polarized intensity above the threshold of 1 mJy. The mean amount of
depolarization can be estimated using the fact that the differential
polarized source counts for unresolved extragalactic sources vary
approximately as $dN/dp \sim p^{-2.5}$ for $1 < p < 10\ \rm mJy$
beam$^{-1}$ \citep{tucci2004,taylor2007}.  If \sigpol\ is a fraction
$f_\Sigma$ of the background value, the polarized intensity per source
must on average be decreased by a factor $f_p = f_\Sigma^{0.4}$.  The
value $f_\Sigma \sim$ 0.5 -- 0.25, typical for the shadows we detect,
corresponds to $f_p \sim$ 0.76 -- 0.57. The polarized intensity of
extragalactic sources in these regions is therefore decreased by up to
$40\%$. The large angular size of the polarization shadows suggests
that these structures are relatively nearby.

Extragalactic sources may be depolarized if $|RM|$ is high enough to
produce differential Faraday rotation over the frequency band
(bandwidth depolarization), or if significant $RM$ fluctuations exist
over the solid angle of the source (beam depolarization).  Complete
bandwidth depolarization for NVSS sources requires $|RM| \sim$ 340
$\rm rad\ m^{-2}$, but Figure 23 in \citet{condon1998} shows that
values of $f_p \sim 0.76$ -- $0.57$ require $|RM| \sim 150$ -- $200$
$\rm rad\ m^{-2}$.  Most polarization shadows are poorly covered by
published rotation measures, but shadows near $l = 90\degr$ that
overlap with the CGPS, coincide with areas where $RM \sim -350$ rad
m$^{-2}$, enough to cause bandwidth depolarization in the NVSS.  The
polarization shadow associated with the Gum nebula comes from a shell
seen in \Halpha\ emission.  \citet{vallee1983} detected this shell
through individual rotation measures $\sim 200$ rad m$^{-2}$,
consistent with partial bandwidth depolarization of NVSS sources by
the shell.

If the depolarization in the NVSS is bandwidth depolarization,
Figure~\ref{shadows-fig} shows regions of high $RM$ up to Galactic
latitude $\pm 20\degr$, with the contour corresponding to $|RM|
\approx 150$ $\rm rad\ m^{-2}$. This $RM$ amplitude map does not
depend on sparsely sampled rotation measures to individual sources,
but it provides complete sampling of the sky north of declination
$-40\degr$, outside the Galactic plane.  We found no evidence for
large-scale structure in an image of $\Sigma_{pol}$ convolved to a
resolution of $5\degr$, providing an upper limit $f_p > 0.94$, or
$|RM| < 70$ $\rm rad\ m^{-2}$ for an extended component outside the
Galactic plane. This upper limit is in general agreement with all-sky
$RM$ images constructed by interpolation of rotation measures of
extragalactic sources \citep{frick2001,JH2004,dineen2005},
although these images do not show structure on angular scales
$\lesssim 20\degr$, because of the limited sky density of published
rotation measures.

Most polarization shadows appear around $l \approx \pm 90\degr$ where
the line of sight is approximately along the local Galactic magnetic
field. Larger values of $|RM|$ are expected here, that can increase
the amount of bandwidth depolarization, but other depolarization
mechanisms may also be enhanced.

Beam depolarization of extragalactic sources implies $RM$ fluctuations
over the solid angle of the source.  \citet{gaensler2005} and
\citet{haverkorn2006} report beam depolarization of extragalactic
sources behind the LMC and in the SGPS. Beam depolarization is also a
possible mechanism for at least some of the polarization shadows
observed here.  Polarized sources fainter than 100 mJy in total
intensity provide most of the signal in Figure~\ref{shadows-fig}a.
Beam depolarization of these sources would require $RM$ fluctuations
on angular scales less than their median angular size of $1\arcsec$
\citep{windhorst2003}. The present data do not allow us to identify
shadows with low $|RM|$ where bandwidth depolarization can be
excluded.  More rotation measures of background sources in
polarization shadows are needed to resolve this question.  If beam
depolarization is significant for compact sources, it must also be
significant for diffuse emission. Polarization shadows caused by beam
depolarization should therefore provide valuable information for the
interpretation of diffuse polarized Galactic emission.

While some deep extended polarization shadows are located in regions
with low \Halpha\ intensity, Table~\ref{shadows-tab} also lists
shadows associated with \HII\ regions where the \Halpha\ emission is a
factor $\sim 10$ brighter. Inspection of Figure~\ref{shadows-fig}
shows that some bright \Halpha\ emission does not give rise to a
polarization shadow, e.g. the \HII\ region NGC 1499 at $(l,b)=$
$(160\degr,-13\degr)$, and the Orion region at $(l,b)\approx$
$(-150\degr,-15\degr)$. Some bright \Halpha\ emission from the Gum
nebula at negative latitudes also does not give rise to a polarization
shadow.  

In the absence of extinction, \Halpha\ intensity in R is related to
emission measure $EM$ in cm$^{-6}$ pc following
\begin{equation}
I_{H\alpha} = 2.25 \int_{\infty}^{observer} n_e^2 dl = 2.25 EM
\end{equation}
\citep{haffner2003} for gas at a temperature of 8000 K.  Regardless of
the depolarization mechanism, $RM$ structure caused exclusively by
variations in electron density, would imply a correlation between
emission measure and the depth of polarization shadows. The lack of
correlation between $\Sigma_{pol,min}$ and $I_{H\alpha}$ in
Table~\ref{shadows-tab}, and the observation that some high $EM$
objects do not give rise to polarization shadows, suggests that
structure in the magnetic field is important in the formation of some
or all polarization shadows.  If all polarization shadows were caused
by bandwidth depolarization in the NVSS, the shadows may be regions in
the interstellar medium where the magnetic field is more regular and
aligned with the large-scale magnetic field than elsewhere. If all
polarization shadows were caused by beam depolarization, they would
represent a more random magnetic field than elsewhere. Both beam
depolarization and bandwidth depolarization may occur, but in either
case polarization shadows suggest a different magnetic structure from
their surroundings.

The size and $|RM|$ of the polarization shadows are of the same order
of magnitude as the outer outer scale ($4\degr - 5\degr$) and standard
deviation ($263\pm 32$ rad m$^{-2}$) of $RM$ fluctuations in interarm
regions in the SGPS reported by \citet{haverkornetal2006a}. The
polarization shadows may be nearby examples of structures
corresponding to this outer scale of $RM$ fluctuations.

\section{Conclusions}

We demonstrate that nearby structures in the magneto-ionic medium
create polarization shadows observed as depressions in the sky density
of polarized sources in the NVSS. Some polarization shadows are
associated with \HII\ regions, but other shadows are related to
depolarized areas in diffuse Galactic radio emission. Partial
bandwidth depolarization in the NVSS is responsible for at least some
polarization shadows, suggesting that these shadows are local regions
of high $|RM|$.  Such regions may affect estimates of the scale height
of the Galactic magnetic field from $RM$ data. If beam
depolarization dominates, polarization shadows constrain
depolarization mechanisms for diffuse polarized emission.

\section*{Acknowledgements}
This Letter is based entirely on published surveys, and was made
possible by the combined effort of the people who made them publicly 
available. The National Radio Astronomy Observatory is a facility of
the National Science Foundation operated under cooperative agreement
by Associated Universities, Inc.  The Virginia Tech Spectral-Line
Survey (VTSS), the Southern H-Alpha Sky Survey Atlas (SHASSA), and the
Wisconsin H-Alpha Mapper (WHAM) are supported by the National Science
Foundation. The Dominion Radio Astrophysical Observatory is operated
as a national Facility by the National Research Council Canada. The
authors thank Jo-Anne Brown for use of her extensive database of
published and unpublished rotation measures, and T. L. Landecker for
his comments on the manuscript.

\clearpage

\begin{deluxetable}{rrrclrr}
\tablecolumns{7}
\tablewidth{0pc} 
\tablecaption{ Polarization Shadows \label{shadows-tab} }
\tablehead{
\colhead{$l$} & \colhead{$b$} & \colhead{a $\times$ b}&\colhead{$\Sigma_{pol,min}$} &\colhead{$I_{H\alpha}$}&\colhead{ID}&\colhead{d} \\
\colhead{($\degr$)} & \colhead{($\degr$)}  &\colhead{($\degr \times \degr$)}&\colhead{(deg$^{-2}$)}& R & \colhead{\ } &\colhead{(kpc)} 
}
\startdata
 $-$125  & $-$7  & \phantom{0}7$\times$\phantom{0}6& 1.5 & 38 & RCW 15\tablenotemark{a} & 1.5\phantom{0} \\
 $-$106  & +4  & 32$\times$17                      & 1.2 & 50 & Gum\tablenotemark{b} & 0.4\phantom{0} \\
  5    & +24 & \phantom{0}5$\times$\phantom{0}3    & 2.7 & 85 & $\zeta$ Oph\tablenotemark{c}  &  0.14 \\
  28   & $-$3  & \phantom{0}6$\times$\phantom{0}4  & 2.6 & 20 & Sct cl.\tablenotemark{d} & $\gtrsim$6\phantom{.} \\  
  47   & +5  & 14$\times$\phantom{0}6              & 1.3 & \phantom{0}3 & \ldots\tablenotemark{e} &\ldots \\
  64   & $-$6  & 14$\times$\phantom{0}7            & 1.5 & \phantom{0}6 & \ldots\tablenotemark{e} &\ldots \\
  93  &  $-$6  & 16$\times$\phantom{0}4            & 1.5 & 10 & \ldots\tablenotemark{f} &\ldots \\
 101  &  +3    & \phantom{0}4$\times$\phantom{0}4  & 2.2 & 78 & \ldots\tablenotemark{g} &\ldots \\
 105  &  $-$2  & \phantom{0}5$\times$\phantom{0}4  & 2.5 & 27 & \ldots\tablenotemark{g} &\ldots \\
 145  & +15  & \phantom{0}3$\times$\phantom{0}2    & 3.3 & 19 & Sivan 3\tablenotemark{h} & 1.2\phantom{0} \\
\enddata
\tablenotetext{\ }{Notes: a. Distance by association with open cluster NGC 2362; \citet{chanot1983}; b. Position is the estimated center of the ring; dimensions are the major and minor axes. \citet{brandt1971}, \citet{chanot1983}; c. \citet{wood2005}; d. Size and location correspond with Scutum cloud described by \citet{madsen2005}; e. Associated with depolarized region in Figure~\ref{shadows-fig}c; f. Comparison with diffuse polarized emission in Figure~\ref{shadows_detail-fig}; g. No clear association with nearby \HII\ regions; h. Distance by association with $\alpha$ Cham; \citet{madsen2006}, \citet{markova2002}}
\end{deluxetable}

\clearpage

\begin{figure}
\caption{ {\bf [This Figure is provided as a separate image f1.gif]}
(a): Surface density of polarized sources in the NVSS with polarized
intensity $>1$ mJy, smoothed to a resolution of 2 degrees.  Gray
scales are linear from 0 (black) to 9 sources deg$^{-2}$ (white).
(b): \Halpha\ intensity from \citet{finkbeiner2003}. Gray scales are
logarithmic from 1 R (white) to 250 R (black). The white contour
indicates \Halpha\ intensity of 25 R.  The southern declination limit
of the NVSS is indicated by the black arc. (c): Polarized intensity of
the Galactic diffuse emission at 1.4 GHz from the DRAO low-resolution
polarization survey \citep{wolleben2006}. Gray scales are linear from
0 (black) to 350 mK (white).
\label{shadows-fig}
}  
\end{figure}

\clearpage

\begin{figure}
\caption{ {\bf [This Figure is provided as a separate image f2.gif]}
(a): Distribution of NVSS sources with $p > 1$ mJy/beam around
$(l,b)=(90\degr,-10\degr)$.  The size of the symbols indicates the
magnitude of $p$, and symbol sizes corresponding with 1, 3, 5, and 10
or more mJy/beam are shown in the inset.  (b) Correlation of \sigpol\
with structure in diffuse Galactic polarized emission in the survey by
\citet{wolleben2006}, for the area shown in (a).  The image shows
\sigpol\ with linear gray scale from 0 (black) to 8 (white) sources
deg$^{-2}$, and contours at 2.7 and 4 sources deg$^{-2}$ (3$\sigma$
and 2$\sigma$ below the mean density of 6.6 sources
deg$^{-2}$). Polarization vectors of diffuse polarized emission are
shown with size proportional to the diffuse polarized intensity. (c):
Image of diffuse polarized intensity with the same polarization
vectors. The region of high polarized intensity and regular
polarization angle crosses between the two minima in \sigpol\ in (b).
\label{shadows_detail-fig}
}  
\end{figure}

\end{document}